\documentclass[10pt, conference, compsocconf]{IEEEtran}

\usepackage{graphicx}
\usepackage{color}
\usepackage{amssymb}
\usepackage{amsmath}
\usepackage{latexsym}
\usepackage{stmaryrd}
\usepackage{alltt}
\usepackage{listings}
\usepackage{url}
\usepackage[utf8]{inputenc}

\usepackage{tweaklist}

\graphicspath{{./figures/}}

{\begin{footnotesize}\begin{alltt}}%
{\end{alltt}\end{footnotesize}}

        \usepackage{epic,eepic,amsfonts,amssymb, amsmath}
        \usepackage{latexsym}

\begin{document}
\date{}

\author{Steffen Bartsch \qquad Karsten Sohr \qquad Michaela Bunke \qquad Oliver Hofrichter
\qquad Bernhard Berger\\TZI, Universität Bremen,
Germany\\\{sbartsch,sohr,mbunke,olihof,berber\}@tzi.org}

\title{The Transitivity of Trust Problem in the Interaction of Android Applications}

\maketitle

\begin{abstract}
Mobile phones have developed into complex platforms with large numbers of installed
applications and a wide range of sensitive data.  Application security policies limit
the permissions of each installed application.  As applications may interact, 
restricting single applications may create a false sense of security for the end users
while data may still leave the mobile phone through other applications.  Instead, the
information flow needs to be policed for the composite system of applications in a
transparent and usable manner. In this paper, we propose to employ static analysis based on the 
software architecture and focused data flow analysis to scalably detect information
flows between components. Specifically, we aim to reveal transitivity of trust problems in multi-component mobile platforms. We demonstrate
the feasibility of our approach with Android applications, although the generalization of the analysis to similar
composition-based architectures, such as Service-oriented Architecture, can also be explored in the future.

\end{abstract}



\fussy

\section{Introduction}
\label{sec:introduction}

Powerful and well-connected smartphones are becoming increasingly common with the
availability of affordable devices and data plans.  Increasingly, the smartphones'
features are provided by focused applications that users can easily install from
application market places.  On the other hand, with tens of thousands of applications
available, there is only limited control over the quality and intent of those
applications.  Mobile code and extensibility is one of the key issues that increase
the complexity of information security \cite{McGraw06}.  To counter this threat,
mobile operating systems impose security restrictions for each application.

\begin{figure}
\centering
\includegraphics[width=0.9\columnwidth]{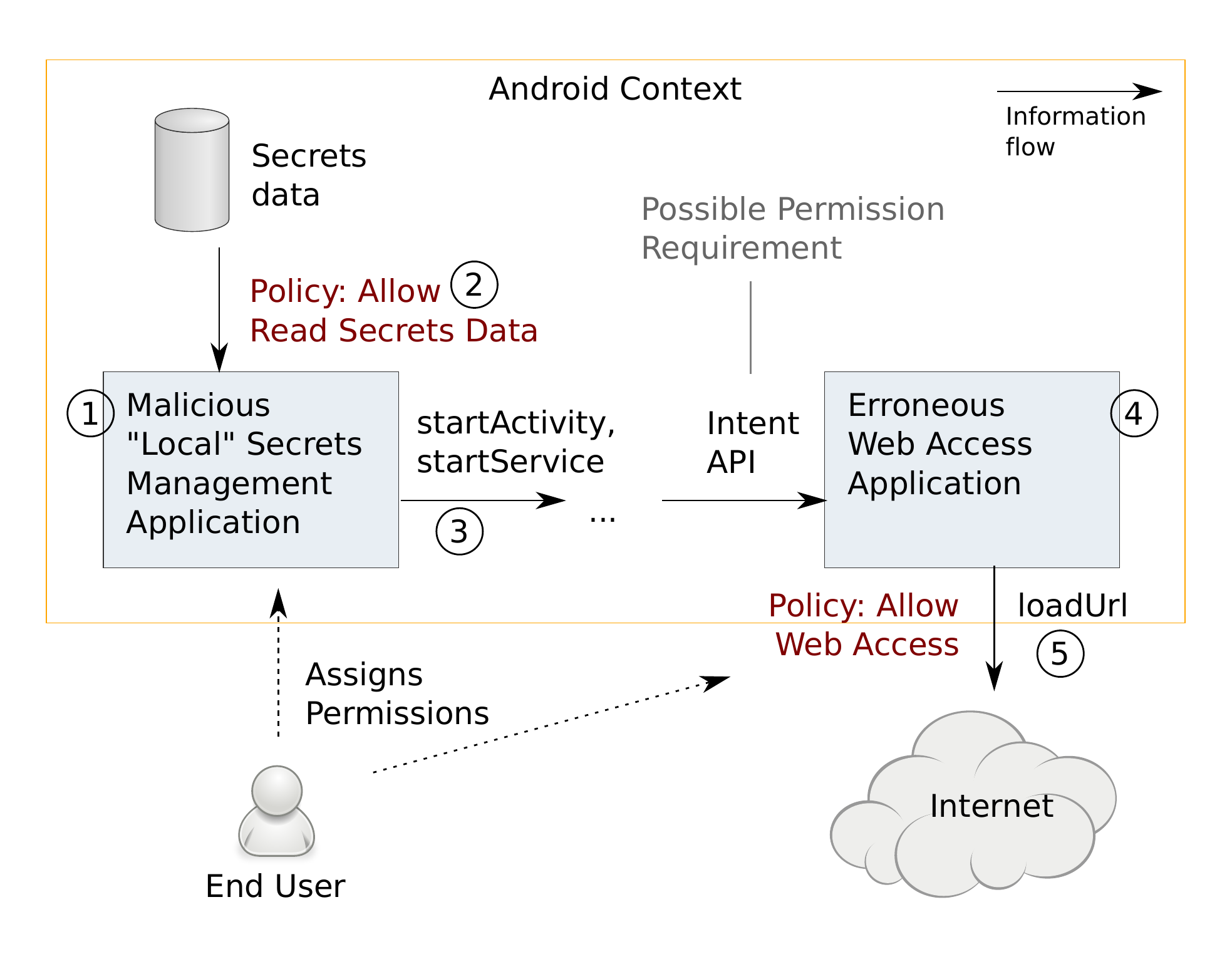}
\caption{Example of malicious information flow}
\label{fig:android-example}
\end{figure}

The Android mobile operating system is one of the major systems on mobile phones.
In case of the Android security model, the least-privilege principle is
enforced through application-level permissions that can be requested by the
applications.  End users need to grant the permissions at install time and thereby decide on the
adequacy of the required permissions and the trustworthiness of the individual
application. The permission granting procedure places a burden on the end users,
who need to reason about how the application might employ the permissions.
In particular, the end user has little knowledge about the consequences regarding the
transitivity of permission granting. As depicted in Figure~\ref{fig:android-example}, a malicious
application (1) with only local permissions (2) may proxy sensitive data (3) through third
party applications and services (4) to external destinations (5).  We describe further attack
scenarios in Section~\ref{sec:android-attack-scenarios}. The
inter-component cooperation is an important concept on the Android platform, but the user
needs to be able to differentiate between legitimate and malicious information flows.

The above-described issue of missing transparency poses a risk with the
spreading of smartphones, the large numbers of available applications and the
prevalent custom of installing applications from untrustworthy sources. The high
likelihood of the threat can be deduced from the wealth of sensitive data that is
stored on mobile phones, ranging from online banking and business application
credentials to communication data and location information. On the Android platform,
the first attacks have already been conducted through malicious online banking
applications \cite{He10}. The threat is further increased by the number of data
channels, such as the short message service, E-mail or Web access that allow the flow of information out of the device context.  Moreover,
attack scenarios are not limited to confidentiality breaches. The integrity of the
device may also be endangered through similar attack vectors. For example,
permissions to use expensive services may be abused.

In this paper, we describe an approach to detecting
illegitimate information flows between different applications and out of the platform. This way,
problems can be revealed that are induced by interacting applications and permission transitivity. We demonstrate the feasibility
of our approach with applications running on the Android platform, although
other mobile platforms and application markets (iPhone Store, Blackberry World,
Windows Mobile Market) are similarly threatened. The transitivity-of-trust problem is not only restricted to
mobile platforms. A similar threat can be
seen in other multi-component environments, such as Social Networking applications
and Service-oriented Architecture (SOA). In this sense, we see our work as a starting point for 
research, namely, analyzing the consequences of the extensibility of systems with
respect to security and mitigating possible risks induced by this paradigm.

Our approach to the information flow analysis spanning multiple applications is as follows. The information on data
sources such as location services, databases, or contact lists are combined with
information on the data sinks (outgoing channels). These input data are used in a
two-layer information flow analysis: First, we identify Android components and the
respective inter-process communication (IPC) points with the help of the reverse engineering tool-suite Bauhaus
\cite{conf/adaEurope/RazaVP06}.  This part of the analysis is carried out at the level of the
software architecture, reducing the analysis effort. In the second step, we use these architectural 
information to slice the code and conduct focused data flow analyses on the software slices,
resulting in the actual information flows that are used to construct an information
flow graph at the architectural layer.  The information flow graph can then be used
by developers and security experts to identify malicious flows and the graph can be
checked against a policy of legitimate information flows.  Moreover, an abstract
representation can help end users in assessing the risk from a new application.
The advantage of the proposed two-layer approach is its scalability. 
In addition, the approach is practically
relevant and has real-world environments as the benchmark. 

In summary, our analysis method can be considered complementary to other static code analysis approaches 
that can dectect implementation bugs such as SQL injection and Cross-site scripting vulnerablities \cite{Livshits05a}. 
Our approach, however, is focused more on the aspect of program comprehension for security and makes transparent interactions between different applications. 


The remainder of this paper is structured as follows. In
Section~\ref{sec:background}, we briefly describe the background on Android,
Android's security concepts and the software analysis tools used for our analyses.
We, then, list possible attack scenarios on the Android platform and show the
relevance of the transitivity of trust problem before discussing the data sources and
sinks of Android applications.  In Section~\ref{sec:approach}, we present our
approach to the security analysis of Android applications in detail, followed by a
case study in Section~\ref{sec:eval}. We discuss the advantages but also limitations
of our approach in Section~\ref{sec:discuss}. After listing the related work, we
conclude in Section~\ref{sec:summary}.

\section{Background}
\label{sec:background}
\subsection{The Android Programming Model}
\label{sec:android-programming-model}

Applications on the Android platform are developed using the Java programming
language. Android applications are not executed on traditional Java Virtual
Machines, but are converted into the custom DEX bytecode and interpreted with the Dalvik
virtual machine. The Android SDK supports most of the Java Platform, Standard Edition
and contains, in addition, Android-specific extensions, including
telephony functions and a UI framework.

Android applications consist of four basic component types: activities, services,
content providers, and broadcast receivers. Activities constitute the presentation layer of
an application and allow users to directly interact with the application.  Services
represent background processes without a user interface. Content providers
are data stores that allow developers to share databases across application
boundaries. Broadcast receivers receive and react to broadcast
messages, for example the ``battery low'' message from the Android OS or messages from other
applications.  For communication between the individual components of applications,
inter-process communication (IPC) provides a means to pass messages between different
applications, activities, and services \cite{Enck2009}.  Android uses messages
that contain meta information and arbitrary data, called intents, for IPC.

Android components follow a lifecycle that is managed by the OS. As a consequence,
there is no \verb+main()+ method from which the applications are started. Instead, the Android OS
calls the lifecycle methods, such as \verb+onCreate()+, whenever e.g.\ an activity is
started for the first time or a new message is received by a service. Further information about
component lifecycles is available from the Android Developer's Guide \cite{Android10b}.

\subsection{Android Security Concepts}\label{sec:AndroidSec}
Android has two basic methods of security enforcement \cite{Enck2009}. Firstly,
applications run as Linux processes with individual Unix users and thus are separated
from each other. This way, a security hole in one application does not affect other
applications. However, as mentioned above, Android provides IPC mechanisms that need
to be secured.  The Android middleware implements a reference monitor to mediate
the access to application components based upon permission labels defined for the
component being accessed.  If an application intends to access a component of another
application, the end user has to grant an appropriate permission. The requested
permissions are specified in the application's policy file.

All permissions requested by an application are granted by the end user at
installation time, i.e., the permission assignment cannot be changed at runtime.
During the installation process, a list of dangerous permissions is presented to
the end user in a dialog window and needs to be confirmed. 

Furthermore, the security model has several refinements that increase the model's
complexity.  One example is the concept of shared user IDs that allow different
applications to share the same user ID if the applications are issued by the same
developer. Another refinement are protected APIs: Several security-critical system
resources can be accessed directly rather than using components. Examples of such
resources are Internet services that allow an application to open arbitrary network
sockets and have full access to the  Internet and outgoing call APIs that allow an
application to monitor, modify, or abort outgoing calls.  In order to mediate access
to such resources, additional in-code security checks have been implemented.
Moreover, permissions are assigned protection levels such as ``normal'', ``dangerous'', and
``signature''. The Android security model also supports delegation concepts such as
pending intents and URI permissions that can only be
checked at the code level rather than at the policy level  \cite{Enck2009}.

\subsection{Architecture-Based Analysis with the Bauhaus Tool}\label{sec:Bauhaus}

The Bauhaus tool-suite is a reverse engineering tool-suite that has been developed
for more than ten years and has been used in several industry projects
\cite{conf/adaEurope/RazaVP06}. Bauhaus allows one to deduce two abstractions from
the source code, namely the Intermediate Language (IML) and the resource flow graph
(RFG).  The IML representation is, in essence, an attributed syntax tree (an
enhanced AST) that contains the detailed program structure information such as loop
statements, variable definitions and name bindings. The RFG representation works
at a higher abstraction level and represents architecturally relevant information of
the software. The RFG is a hierarchical graph that consists of typed nodes and edges
representing elements like types, components and routines and their relations. The
information that is stored in the RFG is structured in views, where each view represents a
different aspect of the architecture, e.g.\ a call graph. Technically, views are
subgraphs of the RFG.

Bauhaus supports a meta-model and thus allows one to define new node and edge types. 
Currently, RFG profiles exist for C/C++, C\#,
and Java, representing syntactical elements of the respective language and their
relations. For example, typical node types for Java are {\tt Class}, {\tt Method},
and {\tt Member}; edge types are {\tt Member Set}, {\tt Member Use}, and {\tt
Dispatching\_Call} among others.  In particular, an extension of the Java-based RFG
model with Android-specific node and edge types is also possible.  

For visualizing the different views of RFGs, the Graphical Visualiser (Gravis) has
been implemented \cite{CEKKS00}. Gravis facilitates high-level analysis of
the system and provides a rich functionality to produce new views by RFG analyses or
to manipulate generated views.

\subsection{Low-Level Analysis with the Soot Tool}

The Bauhaus RFG represents the software architecture, but this abstract
representation lacks detailed program information that is needed for data
flow analysis.  Thus, for our goal of tracing the data flow through the
application, we need an enhanced AST.  The Bauhaus IML-generator
supports Java program code below version 1.5, but for developing and analyzing Android
applications, we need to analyze Java 1.5 code and above \cite{Android10}.
To deal with this issue, we chose Soot, a well-established Java analysis tool
\cite{Vallee-Rai99}, as a basis for performing the data flow analysis.

Soot was designed as a Java bytecode optimization framework in 1999. In the following
years, this framework has been enhanced with several other analysis methods, like
points-to analysis \cite{lhotak03} or dynamic inter-procedural analysis
\cite{Feng04}. Soot provides the ability to inject self-written analyses into the
existing analysis chain on every intermediate representation \cite{Vallee-Rai99}.  Our
analysis takes place on the Jimple representation, a 3-address code representation.
The built-in call graph generation and flow analysis framework
does not facilitate our analysis, since, to take advantage of the Android framework
semantics as described in Section~\ref{sec:android-programming-model}, a custom
static data flow analysis is required.
\section{The Transitivity of Trust Problem}
\label{sec:trans}
We first describe different attack scenarios that may lead to undesirable
information flows on the Android platform. In particular, these scenarios show that
real attacks are possible that exploit transitive trust issues. Thereafter, we argue that the transitivity problem is more general, not restricted to Android or other smartphone platforms. 

\subsection{Threats from Android Applications}
\label{sec:android-attack-scenarios}

\begin{figure*}
\centering
\includegraphics[width=0.9\textwidth]{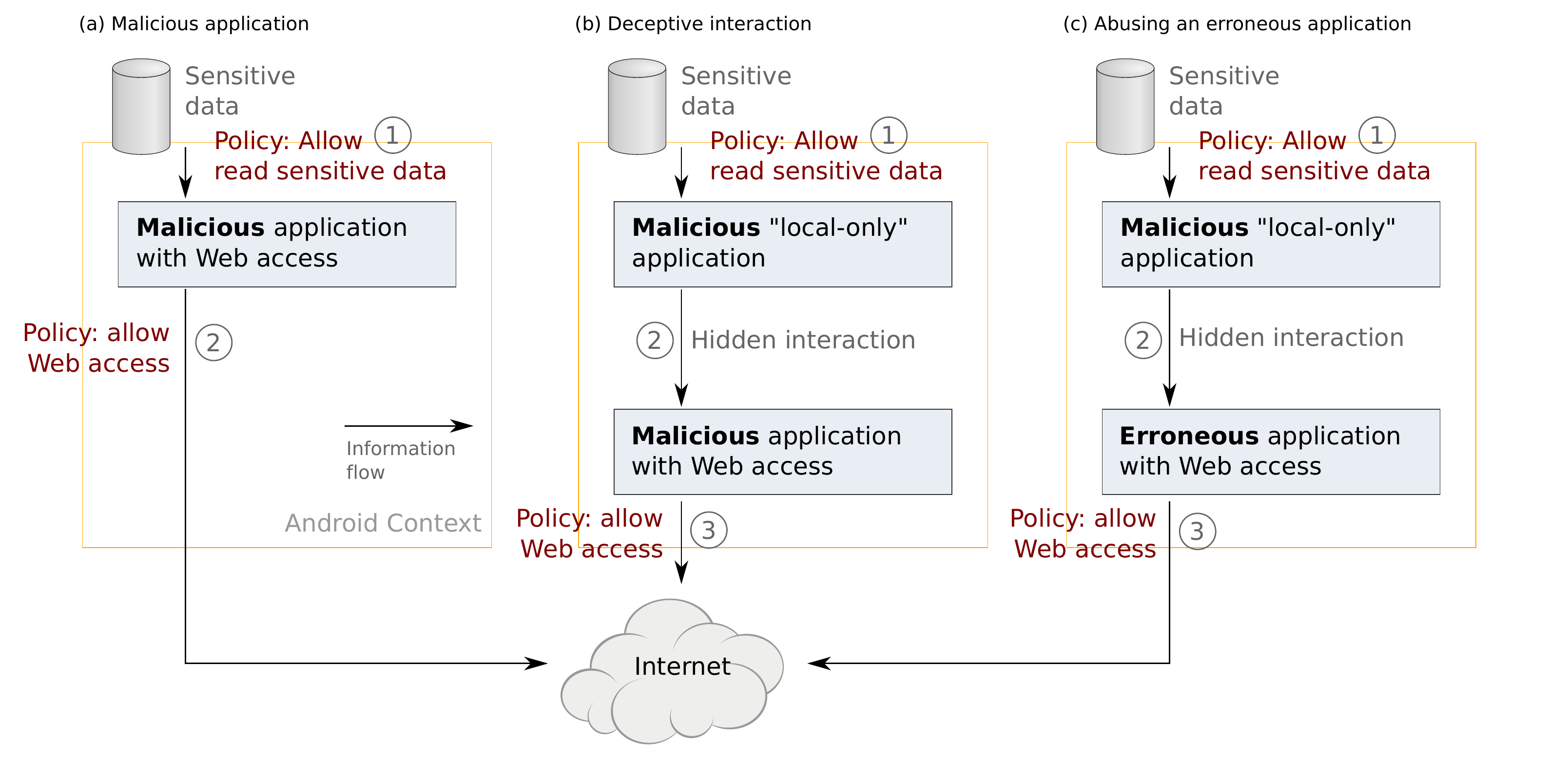}
\caption{Android attack scenarios}
\label{fig:android-attack-scenarios}
\end{figure*}

We identified three classes of attack scenarios through Android applications against
the confidentiality of user data, depicted in
Figure~\ref{fig:android-attack-scenarios}.  In the simplest case, scenario (a), a
maliciously crafted application is published through the Android market.  While the
application may provide a useful function on the surface, behind the back of the
user, it transfers sensitive data (1) to a Web service on the Internet (2).  The Android
security concept requires the end user to notice the combination of permissions to
read sensitive data and access the Internet and cancel the installation.  There are several reasons why this
assumption may fail:
\begin{itemize}
  \item End users are used to accepting permission requests with every
    installation of applications, thus tempted to just acknowledge the shown list;
  \item Many applications require rather broad permissions, for example, Internet
    access for update checks;
  \item The dangerous permissions may be ``hidden'' between less critical or
    irrelevant permissions, such as controlling the display backlight;
  \item In a subconscious risk assessment, the end user may deem the usefulness of the
    application so high that the risk may be acceptable despite the unusual
    combination of permissions.
\end{itemize}
Moreover, as seen in a recent attack, applications can access sensitive data without
explicitly requiring a permission, for example the phone serial number
\cite{Lookout10}.  One way to counter these kinds of attacks is to make the
information flows from critical sources to sinks transparent.  For end users, the permissions
listed in the affirmation dialog could be enriched with indications how these relate
to information flows.

Scenario (b) is considerably more complex since the attacker takes advantage of
application interaction mechanisms in Android.  Similar to scenario (a), the goal is
to compromise the end users' confidentiality by disclosing sensitive data.  To hide
the critical combination of permissions to read sensible data and to send it to
remote services, reading and sending are split into separate applications.  The
first application appears to be ``local-only'' and has read access to sensitive data
(1).  This application interacts using the Android IPC mechanisms, for example, a service binding,
with a second application without the end users' knowledge (2).  The second
application requires Internet access permissions and can thus forward the data to a
remote service (3).  The interaction between the two applications can be completely
hidden from the end user.  There are two approaches to prevent this attack: either to
make the information flows within each application explicit, in this case from a data
source to the Android IPC and from Android IPC to a data sink.  Alternatively,
to analyze all installed applications to identify combinations of applications that can
interact to create an information flow from a critical data source to a remote data
sink.

Scenario (c) is a variant of (b), but with malicious intentions by either the first or the second
application.  If the first one is malicious as shown in
Figure~\ref{fig:android-attack-scenarios}, it will read sensitive data as in scenario
(b).  It will then abuse an erroneous application into tricking it to transfer the
data to a remote target.  In the second case, not shown, an erroneous application
that may read sensitive data offers this data through Android IPC and a malicious
application retrieves the data to forward it to a remote service.  Repositories, such
as OpenIntents\footnote{\url{http://www.openintents.org}}, that offer interfaces for
several inter-component interactions may facilitate this kind of attack.  Apart from
making the critical information flows transparent, it is helpful to validate whether
adequate permission enforcement is enacted on critical information flow paths within
applications, either manifest-based or programmatic to counter this threat.  If
permissions are enforced, end users have a chance of noticing unusual
combinations of permission requirements that do not match with the claimed
application purpose.

\subsection{Transitivity Issues on Different Platforms}
The transitivity of trust problem as discussed before can be regarded as an
instance of a more general security problem in software, namely, undesirable interaction of different applications and components, respectively. This problem has been described in the literature, e.g., by Piessens \cite{Pi02} and Anderson \cite{Anderson2008}.
Transitivity issues between applications are not limited to the Android platform or in general to mobile platforms. Android, however, is a classical example of such systems. First, it supports the mobile code
paradigm which supports dynamically loading new applications.
Furthermore, although there has been implemented a separation mechanism for
applications (or Android components), access between the separated applications is still possible 
via IPC in order to allow the  development of practically useful applications. 

Similar remarks hold for other systems such as multifunction smartcards as stated, for example, by Anderson \cite{Anderson2008}. These cards allow one to install different applications, e.g., one application
for electronic passport functionality and another one for digital signature to allow for legally binding business. In particular, the Javacard technology provides 
the possibility to dynamically load new Java applets \cite{Sun06}. In order to
protect the applications from each other, the concept of ``application firewalls''
has been introduced. Similarly to Android, however, there exists a mechanism to share
Java objects between applications. As a consequence, transitivity issues stemming 
from interacting applications are again conceivable as first discussed by McGraw and Felten \cite{McGFe99}. In this scenario, an application
{\tt A} allows a trusted application {\tt B} to access a sensitive object {\tt x} via a virtual method {\tt x.foo()}. Application {\tt B} then gives 
a third application (not necessarily trusted by {\tt A}) access to a method {\tt y.bar()}, which calls {\tt x.foo()}. This way, {\tt C} indirectly has access to object {\tt x}, although {\tt A} has not explicitly given that permission. 

The interaction problem and specifically transitivity issues also exist in SOA, which
is based on extensible systems such as JavaEE and .NET. In particular, 
Web services, which often implement SOA, aim at coupling and composing services depending on the needs of an organization.
As Karp et al.\ point out, service chaining leads to transitivity problems 
\cite{KarpL10}. We now briefly discuss this point in the context of 
Enterprise Resource Planning (ERP) systems, which often extensively support SOA. 
SAP, for example, makes the NetWeaver platform available to integrate different 
applications such as the Human Resources module or Material Management. 
In particular, SAP NetWeaver can be used to allow different SAP modules or even 
external applications access to centrally administered data such as master data 
(e.g., about customers or vendors), which are sensitive for an organization. 
A business process such as a loan origination workflow, for example, may then 
access or manipulate master data of a customer via a Web service. 
This loan origination workflow itself might be exposed  as a service within the organization. 
If this service is not secured appropriately (e.g., the role-based access control policy is erroneous), 
then it is conceivable that these sensitive data might be accessed by unauthorized actors through the
service.

In summary, application interaction w.r.t.\ transitivity is a prevalent problem on
many platforms. In this paper, we focus on the Android platform, although the
techniques we use can be applied to other systems as well. Specifically, we then need
to map the system-specific programming concepts to our analysis infrastructure and, e.g., introduce 
specific modeling elements at the RFG level. 

\section{Critical Information Sources\\and Sinks in Android}
\label{sec:critical-sources-sinks}

Our approach to information flow analysis is to analyze inter-component flows from
information sources, such as contact lists, to channels through which information leaves the
device context.  Thus, we must identify communication
mechanisms between components as well as critical incoming and outgoing channels on the
Android platform.  The incoming channels
are referred to as ``data sources'', outgoing channels as ``data sinks''.
We identified a list of inter-component communication mechanisms, sources and sinks
by exploring of the Android application framework and the provided samples.
\begin{table}
\centering
\begin{tabular}[]{p{.35\columnwidth}|p{.55\columnwidth}}
  Description & Exemplary API calls \\
  \hline\hline
  Invoke an \newline \verb+Activity+ by \verb+Intent+ \newline (in the foreground) & \verb+Intent intent = new+ \newline\verb+ Intent(this, Receiver.class);+ \newline \verb+startActivity(intent);+ \\\hline 
  Broadcast messages to \newline registered listeners \newline (one-to-many) & \verb+sendBroadcast(intent);+ \newline \verb+sendStickyBroadcast(intent);+ \newline \verb+sendOrderedBroadcast(intent);+ \\\hline
  Communication with \newline \verb+Service+ \newline (in the background) & \verb+startService();+ \newline \verb+stopService();+ \newline \verb+bindService();+ \\
\end{tabular}
\caption{Inter-component communication mechanisms}
\label{tab:inter-component-communication}
\end{table}
Inter-component communication takes place between the Android component types as
described in Section~\ref{sec:android-programming-model}.
Table~\ref{tab:inter-component-communication} lists the primary 
communication types on the Android platform.  For the sake of brevity, only individual examples of
the API calls are given.

\begin{table}
\centering
\begin{tabular}[]{p{.25\columnwidth}|p{.15\columnwidth}|p{.45\columnwidth}}
  Data source & Criticality & Accessible data \\
  \hline
Content Provider & high & contains passwords, contact list \\
SMS/MMS & high & sensitive information \\
User input & high & passwords \\
Files & high & business documents \\
Network (HTTP) & high & protected files \\
Bluetooth & high & contacts, files, images \\
Camera & medium & observation of image data \\
C2DM & medium & sensitive URI \\
Location Manager & medium & observation of location data \\
Device identifiers & medium & personal identification \\
\end{tabular}
\caption{Data sources (incoming channels)}
\label{tab:sources}
\end{table}

\begin{table}
\centering
\begin{tabular}[]{p{.09\textwidth}|p{.07\textwidth}|p{.25\textwidth}}
  Data sink & Attack\newline complexity & Attack scenario/\newline attack requirements (exemplary) \\
  \hline\hline
Network (WebView) & medium & manipulation of URI/\newline access to URI \\
  \hline
SMS/MMS & medium & manipulation of data or number \\
  \hline
Bluetooth & high & influencing the transferred file/\newline proximity, control of receiver, completed pairing \\
  \hline
Content Provider & high & malicious application misleads into writing into content provider/\newline need to specify content provider URI \\
  \hline
Files & high & malicious application misleads into writing into files/ \newline need to specify file name \\
  \hline
Google Translate API & very high & manipulation of address  resolution/ \newline access to OS services \\
  \hline
MapView & very high & manipulation of address  resolution/ \newline access to OS services \\
\end{tabular}
\caption{Data sinks (outgoing channels)}
\label{tab:sinks}
\end{table}

The origin of the data in a information flow needs
to be known to effectively analyze the flows' criticality.  Table~\ref{tab:sources}
provides a list of data sources that allow the flow of information into the device
and application context.
Enck et al.\ similarly identified data sources for the placement of security hooks in
their dynamic analysis, categorizing sources into sensors, such as location
sensors and camera, information databases and device identifiers
\cite{Enck10}.  
The criticality of a data source is determined by the data
that the source makes potentially accessible.  As the criticality depends on various
factors, we only evaluate the criticality for average users at this point to give a
risk estimation.  We will conduct an in-depth analysis of source criticality as part
of our future work.  For the criticality values in Table~\ref{tab:sources}, high
criticality indicates that the impact is potentially significant. Medium criticality
is assigned for observation scenarios, where consequences resulting from attacks are
limited for average users, depending on the monitored person in a given case. Low
criticality indicates that there is only little impact that most users might accept
the annoyance. An example of a data source with medium criticality is
Android's Location Manager, which provides access to the device's geographical
location and is used in this paper's case study. Bluetooth is a data source with
high criticality because of the possibility to access critical data like contacts,
files and images on paired devices through this channel.

In Table~\ref{tab:sinks}, we list data sinks of Android applications with possible
attack scenarios and the requirements for the realization of this scenario as well as
a valuation of the attack complexity.  While there are severity metrics for software
vulnerabilities, the existing models do not match the requirements of the evaluation of
information flow data sinks.  We assess the attack complexity through the complexity
of possible attack scenarios.  For medium attack complexity, it
is sufficient for a malicious application to trick a single application into proxying
sensitive data to external destinations.  In cases that require several applications
to be coordinated for an attack, we rate the attack complexity as high. Very high
attack complexity indicates that it is, in addition, necessary to modify the operation
system and/or external services, such as the Google Maps Web service.  An example of
a data sink with medium attack complexity is a WebView which displays
web pages as part of the UI. To channel data out of the device context through this sink, a malicious
application has to manipulate the target URI.  In contrast, content providers are
data sinks with high attack complexity because a malicious application must
mislead one application into writing into a content provider and another component
afterwards into using this content provider to channel data out.

Following the standard risk assessment approach of $\mathit{Risk} =
\mathit{Probability} \times \mathit{Impact}$, the risk of an information flow can be
approximated from the source criticality (impact) and the sinks attack complexity (probability).
Thus, a low attack complexity of data sinks combined with a high criticality of data
sources results in a maximum risk.
\vspace{1cm}
\section{Information Flow Analysis of Android Applications}
\label{sec:approach}
To improve the transparency with respect to the transitivity of trust problem on the
Android platform, we propose to analyze the information flows between the applications.
We first introduce the analysis method on a high level before we describe our
prototype implementation of the analysis.

\subsection{Analysis Method}
Our analysis approach aims to identify undesirable information flows between
different Android applications and components, respectively. In order to analyze a
larger set of applications (as it usually exists on an end user's phone), we did
not directly employ the AST for the entire analysis.  We
rather employ two layers
of abstraction in the analysis, beginning at the level of the software architecture
to identify the Android components,
before diving into the AST details to enrich the architecture and, lastly, deriving
the final results from the architecture. In this last step, we employ the architecture to
compose information flows through single Android components into information flows spanning an entire application, and thereafter
compose these intra-application information flows to inter-application information flows. 
All architectural-level analyses are conducted on a
hierarchical architecture graph that represents architectural elements, such as
methods and classes, as nodes and relations between the elements, such as calls, as
edges. In the following, we explain our analysis algorithm in more detail.

\begin{lstlisting}[language=ruby,label=pseudocode,caption=High-level analysis algorithm,float]
# input : applications (applications to be analyzed)
# output : critical_flows (flows that lead from device sources to device sinks and could leak information)

# CodePoint: [Class, Method, Call]
# PointSpec: [PointType, CodePoint]
# Flow: [Component, PointSpec, PointSpec]
inner_application_flows = []
critical_sinks = []
critical_sources = []

applications.each do |application|
  inner_component_flows = []
  source_points_of_entry = []
  sink_points_of_exit = []

  components = application.identify_components
  components.each do |component|
    points_of_entry = component.identify_points_of_entry
    source_points_of_entry += points_of_entry.select_sources
    points_of_exit = component.identify_points_of_exit
    sink_points_of_exit += points_of_exit.select_sinks

    inner_component_flows += inner_component_flows_for(component, points_of_entry, points_of_exit)
  end

  inner_application_flows += inner_application_flows_for(application, inner_component_flows, source_points_of_entry, sink_points_of_exit)

  critical_source += source_points_of_entry.select_device_sources
  critical_sinks += sink_points_of_exit.select_device_sinks
end

critical_flows = []
flow_graph = generate_flow_graph(inner_application_flows)
critical_sinks.each do |critical_sink|
  reachable_elements = flow_graph.reachable_from(critical_sink)
  critical_sources.each do |critical_source|
    if reachable_elements.contains(critical_source)
      critical_flows << Flow.new(nil, critical_sink, critical_source)
    end
  end
end
\end{lstlisting}

A listing of the high-level algorithm is shown in Ruby-style pseudo-code in
Listing~\ref{pseudocode}.  The algorithm starts from a set of Android
applications that should be considered for inter-application information flows.  In
the first architectural analysis phase, Android components are identified for each
application.  Components are the basic entities in the Android programming model that
communicate through IPC, including services, activities and broadcast receivers (see
Section~\ref{sec:android-programming-model}).  In further architecture-level
analyses, we search for the IPC entry and exit points for each component using
architectural patterns.  These points are the basis for the detailed analysis of
intra-component information flows at the AST level.  As shown in
Listing~\ref{pseudocode-component}, backward slicing is conducted on the AST,
starting from each of the exit points.  The goal is to identify information flows
that reach one of the entry points of the component, representing an intra-component
information flow.

\begin{lstlisting}[language=ruby,label=pseudocode-component,caption=Information flows
inside each individual component,float]
def inner_component_flows_for (component, points_of_entry, points_of_exit)
  flows = []
  points_of_exit.each do |point_of_exit|
    backward_slice = point_of_exit.backward_slice_on_ast
    points_of_entry.each do |point_of_entry|
      if backward_slice.contains(point_of_entry)
        flows << Flow.new(component, point_of_entry, point_of_exit, backward_slice)
      end
    end
  end
  return flows
end
\end{lstlisting}

At the architectural level, the intra-component flows are used to enrich the
information flow graph with communication links within each application, resulting in a
component-level flow graph.  Next, we identify information flows on the level of
individual applications.  We focus on flows that originate outside the application
context, pass through it and leave the application again.  As depicted in Listing~\ref{pseudocode-application}, we
conduct a reachability search on the flow graph to find the information flows within each
application.  Searching the flow graph significantly reduces the search space since we only consider
the identified flows and not the entire AST.  We start out from selected entry and
exit points: sources and sinks.  Sources are entry points of components that connect
to the outside of the application, for example, receiving broadcasts.  Sinks are the
opposite, those exit points that leave from the application, for example, starting
application-external activities or accessing Web pages.  From the reachability
analysis on the component-level flow graph, we identify all flows between sources and
sinks within the application.

At this point, we found information flows that pass through applications, but only
for individual applications.  For inter-application flows, in the last phase, we
identify the information flows that involve critical sources and sinks as described in
Section~\ref{sec:critical-sources-sinks}.  An application-level flow graph is constructed from the
intra-application flows as the basis for the identification of critical flows.
Again, a reachability search is conducted, starting from critical sinks and searching for
flows to one of the critical sources.  As a result, all information flows are known
that originate at critical sources and terminate at critical sinks.





\begin{lstlisting}[language=ruby,label=pseudocode-application,caption=Information flows
inside each individual application,float]
def inner_application_flows_for (application, inner_component_flows, source_points_of_entry, sink_points_of_exit)
  flows = []
  flow_graph = generate_flow_graph(inner_component_flows)
  sink_points_of_exit.each do |sink_point_of_exit|
    reachable_elements = flow_graph.reachable_from(sink_point_of_exit)
    source_points_of_entry.each do |source_point_of_entry|
      if reachable_elements.contains(source_point_of_entry)
        flows << Flow.new(application, source_point_of_entry, sink_point_of_exit)
      end
    end
  end
  return flows
end
\end{lstlisting}

\subsection{Prototype Implementation}
\label{sec:prototype}

\begin{figure}
\centering
\includegraphics[width=0.8\columnwidth]{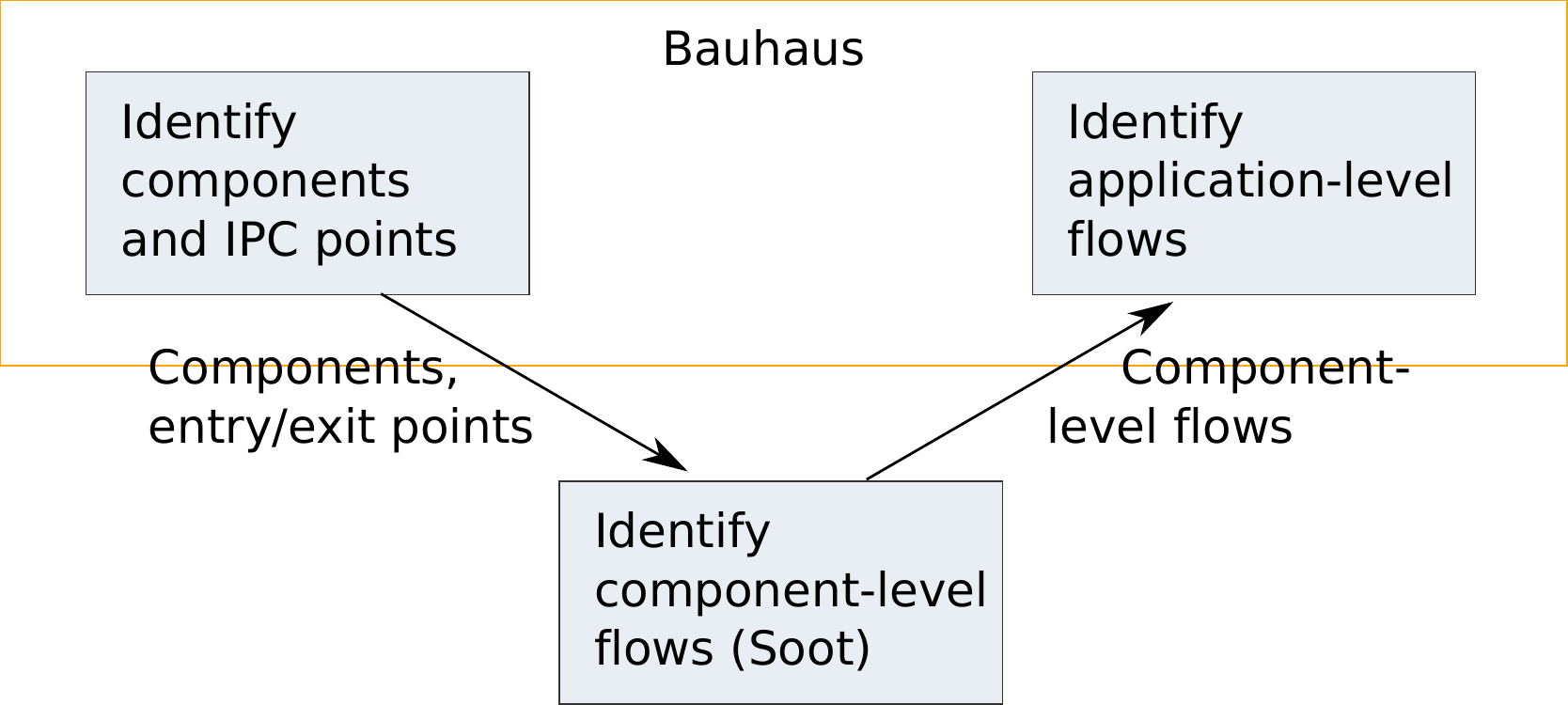}
\caption{Analysis workflow}
\label{fig:analysis-workflow}
\end{figure}

We implemented the  analysis method described above as a prototype that
identifies information flows between Android applications and the Android platform.
The prototype uses two distinct tools to implement the analysis.  As depicted in
Figure~\ref{fig:analysis-workflow}, we employ the Bauhaus tool suite at the
architectural level and the Soot tool for AST-based analyses.

\paragraph{Identify components and IPC points}

In the first phase of the analysis, we identify the components as well as the associated entry
and exit points on the architectural level.  The architecture-level analyses are based upon the 
RFG that the Bauhaus tool generates from the Java bytecode of the analyzed
applications. From the global RFG, we create a reduced information flow subgraph (view),
containing only the relevant parts of the studied components. The relevant
parts are identified through the search for relevant Android framework patterns
that are described in Section~\ref{sec:critical-sources-sinks}.
The Bauhaus tool-suite provides Python language bindings to read and modify
RFGs.  We developed a Python script that prototypically identifies relevant parts through
pattern matching and marks
the related methods, classes and calls by adding the nodes and edges to an
information flow view.

To interface with the further, AST-based data flow analysis, the identified
methods and classes are listed together with the critical calls in an XML-based exchange
format that is passed to the next analysis stage.  

\paragraph{Identify component-level flows}
\label{subsec:intra-comp-data-flow}

In order to find information flows between entry and exit points in components, an
intra-component data flow analysis is carried out at the AST
level.  We use the previously identified entry and exit points to focus the data-flow analysis and
significantly reduce the analysis effort at this level.
We developed analysis algorithms for the Soot tool that utilize the known Android framework semantics.
For each class of entry and exit points that is supported by the prototype, a corresponding analysis
building block has been implemented. 
The behavior of the Android platform prevents the Soot framework from generating 
a sufficient call graph for our analyses.  One reason is that there is no
single entry point to the Android applications, such as the traditional
\texttt{main()} method,
but several, depending on the IPC mechanism.  More importantly, there are
several, partly dynamic framework semantics that need to be part of the call graph, such as UI event
handlers, but are difficult to be statically analyzed.

To identify the intra-component information flows, we search for all program points that affect 
a given exit point in a component.  Therefore, we chose a static backward slicing
technique as described by Weiser \cite{Weiser1981}.
If the backward slicing reaches an entry point of the component under investigation,
we consider this an information flow for the specific entry and exit points.

\paragraph{Identify application-level flows}

The component-level flow
data from the AST-based analysis is now employed to enrich the information flow graph.
The primary purpose of the information flow graph is to allow developers and
security experts to quickly identify flows and determine the risks related
to the flows.
The information flow graph is represented as an RFG view in the Bauhaus tool suite
and is extended as follows.  For each information flow that has been identified in the previous analysis
step, the intra-component flow edge is drawn between the entry and exit point and the
corresponding nodes are added to the view.  If the origin of the
current flow is of type ``source'', an information flow edge is
inserted from the origin to the point of entry inside the current information
flow.  Additionally, for all types of destinations, an edge is inserted between the
point of exit and the destination's point of entry.
We derive the application-level information flows by conducting a reachability
analysis based on the information flow graph, starting from exit points that leave
the application, backwards to entry points that enter the application.  In a last
step, we combine the RFGs of multiple applications to identify the critical flows
between sources and sinks in the application ecosystem again through a reachability
analysis.  The resulting view for the case study below is shown in
Figure~\ref{fig:gravis-analysis-results} as it is displayed in Bauhaus' Gravis tool.


\section{Evaluation}
\label{sec:eval}
We evaluate our approach by means of a case study of a public transport-related
application and thereafter show how the analysis results can be displayed in a usable manner.

\begin{figure}
\centering
\includegraphics[width=0.95\columnwidth]{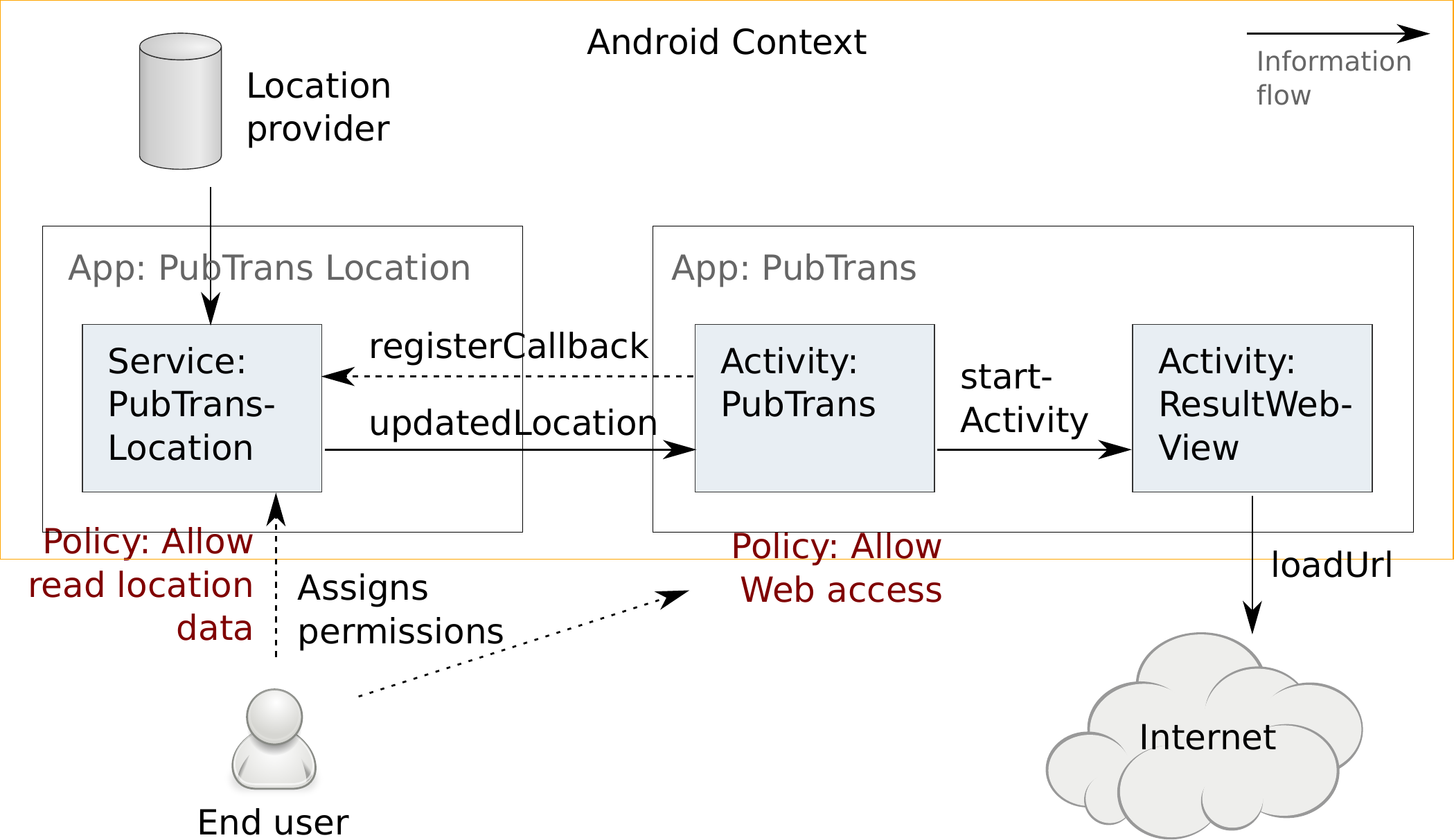}
\caption{Case study setup for public transport application}
\label{fig:efa-query-case-study}
\end{figure}

\subsection{Public Transport Application}

As a case study, we chose to analyze two real-world Android applications that are
available on the Android Market with installations on more than
2000 devices. There are several reasons for selecting these two applications.  One
pragmatic aspect is that one of the authors developed the applications so that we had
access to the applications' bytecode as the basis for our analysis.  The case study
did not affect the design or implementation of the applications since the development
of the applications were finished when our research work on the analysis started.
A more important criterion for choosing this application was that it encompassed
different frequently-used Android concepts such as starting activities or binding to
services and a multi-component design (see Section~\ref{sec:android-programming-model}).  With this case study, we
also demonstrate that our analysis approach supports more complex semantics of the
Android framework such as registering and executing remote service callbacks.  The
two applications also demonstrate that the transitivity of trust is a necessary
concept, although the missing transparency may cause the concept to be misused.

The first application is called ``PubTrans''\footnote{To preserve the anonymity of the
submission, we employ pseudonyms for the applications and removed the original
package names.}, an interface
to a public transport routing Web application, primarily improving the
input form to take advantage of the context (current location and time as well as previous
searches).  PubTrans takes parameters such as origin, destination
and desired arrival time and sends a query to the routing Web
application.  Thus, the PubTrans application requires unrestricted Web access privileges.

When entering the public transport routing parameters, the user may
choose to take the current address as the origin. Since using detailed location
data is not strictly necessary for the application's main goal, location queries
have been factored out into a separate Android component that is installed
as a separate application.  As shown in Figure~\ref{fig:efa-query-case-study}, the
PubTransLocation application thus requires location data permissions.  With two
separate applications, a user may choose whether she would like to grant location
access.  Still, as shown in the figure, there is an information flow between
both applications.  This information flow is required to fulfill the
intended goals, but it was not explicitly granted at installation time.  Although not
harmful, this information flow is an example of the missing transparency with regard
to the transitivity of trust on the Android platform.




\begin{lstlisting}[language=xml,label=lst:exchange-format-example,caption=Excerpt
of analysis input,float]
<component>
  <class>com...pub_trans.ResultWebView</class>
  <point-of-entry type="start-activity">
    <origin type="activity">
    </origin>
    <class>com..pub_trans.ResultWebView</class>
    <method>onCreate</method>
    <call>getIntent</call>
  </point-of-entry>
  <point-of-entry>
    ...
  </point-of-entry>
  <point-of-exit type="call">
    <destination type="sink">
      <class>android.webkit.WebView</class>
    </destination>
    <class>com..pub_trans.ResultWebView</class>
    <method>loadResults</method>
  </point-of-exit>
  <point-of-exit>
    ...
  </point-of-exit>
</component>
<component>
  ...
</component>
\end{lstlisting}

We now describe our analysis approach in more detail with the help of this case
study.  In Figure~\ref{fig:efa-query-case-study}, we can see that our system consists
of two applications with three Android components.  We can identify the entry and
exit points of the components by means of the architecture-level analysis (``Identify
components and IPC points'' step in Section~\ref{sec:prototype}).  Taking a closer
look at the \verb+ResultWebView+ activity, we obtain as an entry point the
\verb+onCreate()+ method and as an exit point the call to the Android \verb+WebView+
UI element, which is at the same time a possible data sink as described in
Section~\ref{sec:critical-sources-sinks}.  Thereafter, we interface with the Soot tool
to perform the detailed analysis on the AST.  We use an XML-based exchange
format to pass on the architectural information to Soot.  The component description
excerpt in Listing~\ref{lst:exchange-format-example} displays two IPC points in the
\verb+ResultWebView+ component.

\begin{figure*}
\centering
\includegraphics[width=.9\textwidth]{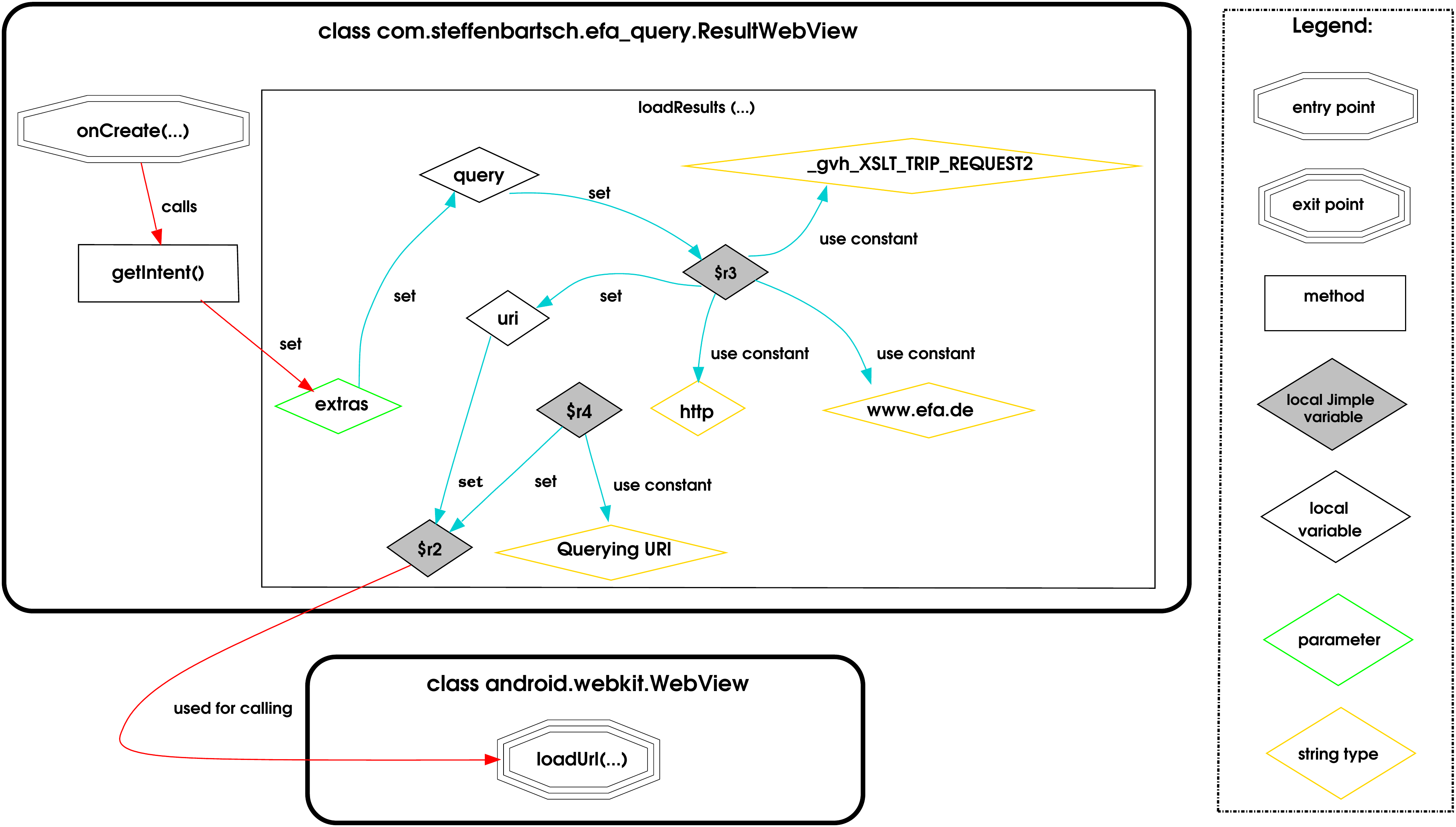}
\caption{Data flow analysis of the ResultWebView component}
\label{fig:analysis-visualized}
\end{figure*}

\begin{figure*}
\centering
\includegraphics[width=0.9\textwidth]{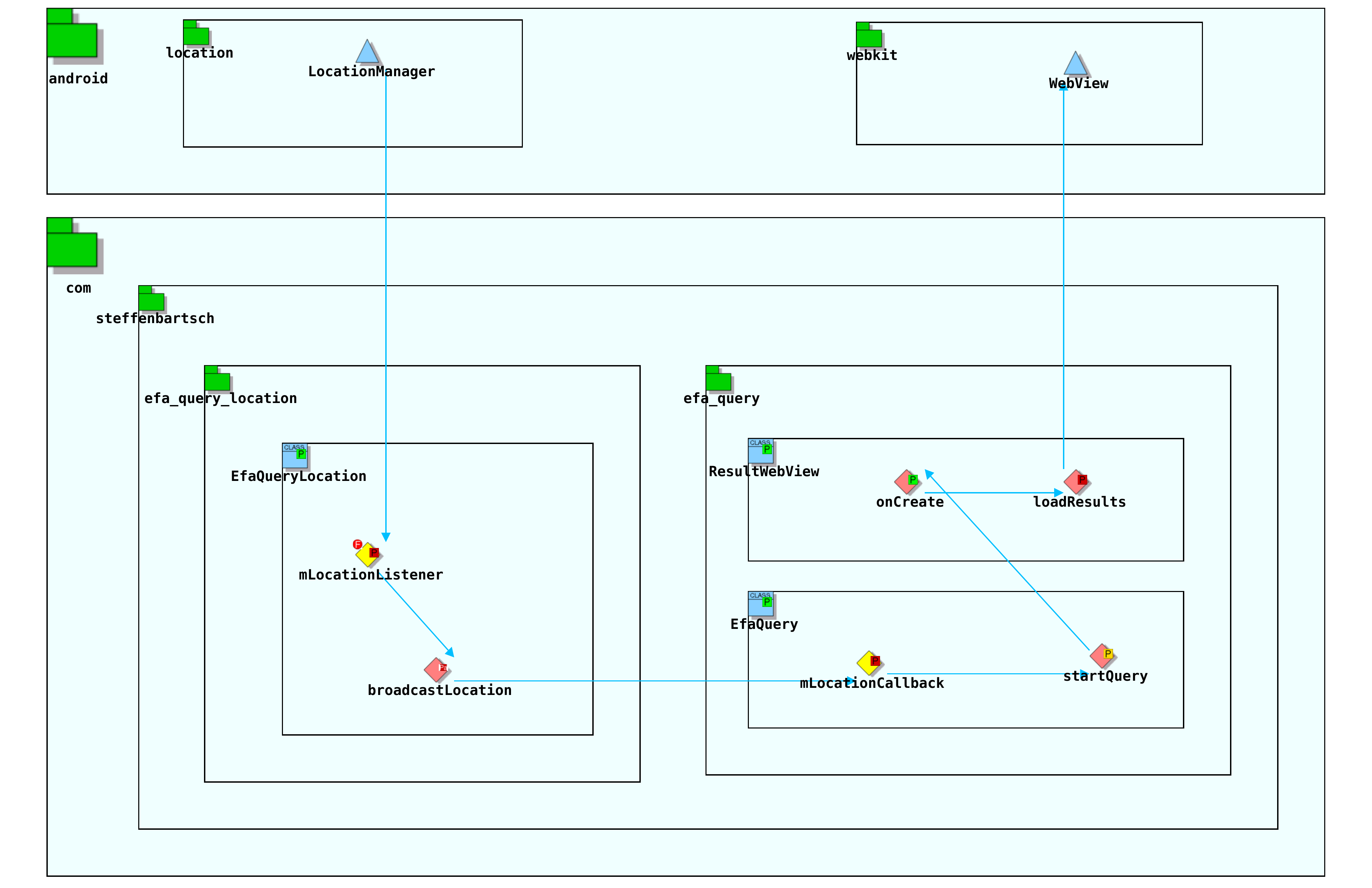}
\caption{Generated information flow view in the Gravis tool}
\label{fig:gravis-analysis-results}
\end{figure*}

On the AST level, we carry out the Soot analysis with the backward slicing algorithm
(``Identify component-level flows'' step), based on the component and IPC point
information.  For the \verb+ResultWebView+ component, we need to verify, for
example, whether an intra-component information flow exists between the IPC points
\verb+onCreate()+ and \verb+WebView+.  Figure~\ref{fig:analysis-visualized} depicts a
didactically simplified excerpt of the backward slice production corresponding to the
information flow from the \verb+ResultWebView+ entry point \verb+onCreate+ to
\verb+WebView+.  In the figure, the backward slice starting point is the exit point
of the component, calling \verb+loadUrl()+ on a \verb+WebView+ component.  Beginning here, we look up all variables passed with the method call and move backward
along the statements inside the method \texttt{load\-Results()} to identify each point
that affects the exit point. When the beginning of this method is reached, we have a
set of variables that affect the exit point and we evaluate whether any of these are
method parameters to do further analysis on affected points, maybe, in other methods
of the \verb+ResultWebView+ class.  In the shown case, the variable \verb+extras+,
marked green in Figure \ref{fig:analysis-visualized}, is a parameter, so we trace
where the current method was called.  The method was called by
\verb+onCreate()+ that was described as a starting point for activities in
Section~\ref{sec:android-programming-model}.  Inside \verb+onCreate()+,
\verb+loadResults()+ is called with the returned value of the inherited method
\verb+getIntent()+.  This inherited method is another artifact of Android activities
and returns the intent with a set of parameters that started the activity.  Thus, we
identified an intra-component information flow between the entry point
\verb+onCreate+ and the exit point \verb+WebView+.


\begin{lstlisting}[language=xml,label=lst:exchange-output-example,caption=Component-level
flows as output from analysis,float]
<information-flow status="checked">
  <class>com...pub_trans.ResultWebView</class>
  <origin type="activity">
    <class>com...pub_trans.EfaQuery</class>
  </origin>
  <point-of-entry type="start-activity">
    <class>com..pub_trans.ResultWebView</class>
    <method>onCreate</method>
    <call>getIntent</call>
  </point-of-entry>
  <point-of-exit type="call">
    <class>com..pub_trans.ResultWebView</class>
    <method>loadResults</method>
  </point-of-exit>
  <destination type="sink">
    <class>android.webkit.WebView</class>
  </destination>
</information-flow>
<infortmation-flow>
  ...
</infortmation-flow>
\end{lstlisting}

The results from the AST-level analysis are passed back to the architectural analysis
through an exchange file, shown in Listing~\ref{lst:exchange-output-example}. In the
last step, the information flow data are used to draw appropriate edges in the
information flow view on the architectural level (``Identify application-level
flows'' step). The information flow view of the resulting RFG is shown in
Figure~\ref{fig:gravis-analysis-results} as displayed in the Bauhaus Gravis
visualization.  When comparing the information flow graph to the case study set up in
Figure~\ref{fig:efa-query-case-study}, one can follow the information flow from the
location provider source through the three components to the Internet.  Thus, the
developed method successfully identified the potentially harmful, at least
intransparent information flow.  While this detailed visualization is helpful for
developers who need to find out which architectural elements are related to
potentially unexpected information flows, security engineers and end users need
significantly more abstract visualizations.  For this reason, further graph searches
are conducted to identify the critical flows that are to be displayed at higher
levels of abstraction.


\subsection{Visualization for Information Flow Transparency}

\begin{figure}
\centering
\includegraphics[width=0.5\columnwidth]{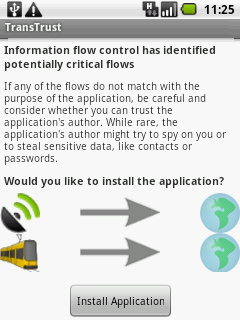}
\caption{End-user information flow visualization}
\label{fig:end-user-ui-mockup}
\end{figure}

As indicated in the previous section, the developer-oriented visualization in
Figure~\ref{fig:gravis-analysis-results} is too detailed to be of use for end users.
To provide an adequate level of abstraction, we generate a more abstract visualization
from the analysis results, depicted in Figure~\ref{fig:end-user-ui-mockup}.  The goal
is to provide insights into the potentially malicious information flows between the
applications and critical sources and sinks.  We display those information flows that
take advantage of the transitivity of trust.  First, we show all information flows
that start out at a critical source and lead to a critical sink.  Additionally, to
prevent false negatives, we also display flows to the sink from applications on the
path.

One option is for end users to employ this visualization
on-demand to gain an overview of hidden information flows on
their devices.  Arguably even more effectively, the visualization might serve as an
addition to the existing installation process.  In this case, additional information
flows that are facilitated by the new application are shown after the user has
accepted the permissions that the application requires but before the actual
installation.  The latter case is what Figure~\ref{fig:end-user-ui-mockup} depicts.

While currently implemented as a separate application, the information flow transparency
view could be integrated into the installer at a later time.  Also, the
flow transparency application currently reads the information flow data from a file
that has been previously generated on a PC. We may port the analysis to the
Android platform as part of our future work.  Alternatively, the analysis may be conducted by the
Android Market owner when the application is uploaded to the market and provided at
installation time in addition to the application package.  The generated information flow data may additionally be used by
security engineers to assess the security of a set of applications.  For example, the
market supervision could use an appropriate visualization to identify potentially
harmful applications.  Information security staff at companies might also be
interested in analyzing the security of applications on their employees' smartphones.

\section{Discussion}
\label{sec:discuss}
We discuss the advantages as well as the limitations of our approach. In particular,
this discussion covers aspects such as false positives, scalability and the relationship to dynamic analysis.

\subsection{Case Study Results}
The positive result from the case study is promising, in
particular, when considering that the studied applications are real-world applications
that have not been modified for the developed method.  However, the proposed method
depends on framework semantics and thus is limited to those sources, sinks and IPC
mechanisms that are implemented.
As of now, we have discovered the critical mechanisms
that are listed in Section~\ref{sec:critical-sources-sinks}.  Currently, the
algorithm is implemented for the location provider source, the \verb|WebView| sink as
well as \verb|startActivity| and \verb|bindService| IPC.  We are planning to add the other
mechanisms as part of our future work. In general, unsupported source or sink types will result in 
false negatives.  Moreover, while evaluating the
proposed method, we discovered two aspects that are difficult to automatically
analyze and may lead to additional false negatives.  Firstly, user inputs can contain credentials and need to be taken into
account as a data source.  Also, information flows involving content providers as
intermediate step are difficult to follow without deeper knowledge of content
provider URI semantics.

\subsection{False Positives}

Similar to other static analyses, our approach may lead to false positives that
can be produced on different levels of our analyses. The
first source of false positives is that we find all 
information flows from critical data sources to data sinks irrespective of whether
they are intended or unintended.
In our case study, it is intended by the programmer that the location information
is passed to the Web page because it is the only way to retrieve the requested
information.  Therefore, this information flow is a false positive from the programmer's
viewpoint. And it may also be a false positive for an end user if she is aware that
the feature necessarily needs the location data.

The second source of false positives is the underlying data flow analysis. This
analysis may find data flows that are not existent due to overestimation. In
such a case, we may identify a connection between a data source and a data sink
within an application that may not occur in practice. This false positive on the lower level
of our analysis may lead to false positives in the set of possible inter-component
information flows.


\subsection{Scalability}

The highest complexity of the proposed algorithm lies in the backward slicing
algorithm. Specifically, the construction of the internal data structure for slicing,
i.e., the AST or to be more precise, the program dependence graph (PDG) is the
time-consuming step \cite{BiHoRe90}. By isolating the backslicing runs for each
component, we are confident that the whole algorithm will scale well in relation to
the number of examined components.  By selecting only potential paths between the
components' entry and exit points with the help of the RFG (see the step ``Identify
components and IPC points''), the number of the nodes of the PDG will be reduced. We
do not have to build a complete PDG for all Android components. The effort for the backward slicing
algorithm then is linear in the size of the PDG \cite{BiHoRe90}.

\subsection{Dynamic versus Static Analysis}

The focus of our work lies on static code analysis, which in principle can be carried
out offline, e.g., on other hardware as done in our current prototype implementation.
Dynamic analysis is another approach to address the problem of undesirable
information flows on Android.  Specifically, the TaintDroid tool implements dynamic
monitoring of privacy-relevant flows by modifying the Dalvik VM and the Android
kernel \cite{Enck10}. Instead of static analysis before installation, TaintDroid
complements our approach by offering analyses at runtime.  While TaintDroid aims to
minimize the performance overhead, static analyses can also afford to carry out more
detailed analyses. For example, the tools employed by our approach allow us to even
detect indirect information flows induced by control flows
\cite{conf/acsac/ChandraF07} although this is not the topic of this paper. 

One argument in favor of the dynamic analysis of Android applications is that no
source code is needed \cite{Enck10}.  Android uses a different
distribution format called DEX, which is a customized bytecode format being
register-based rather than employing an operand stack. However, we have obtained
promising early results with the dex2jar tool, which translates DEX to Java bytecode
\cite{Dex10}. Since our analyses work on Java bytecode (see
Section~\ref{sec:prototype}), we also applied our tools to DEX code for the case
study.  However, applying the dex2jar tool to larger application sets remains future
work. 

Not all properties can be inferred from the code statically. One example is the
implicit intent decomposition mechanism in the Android framework. It decides at
runtime among all registered components which component offers the requested intent
features and accordingly suits to the implicit intent call. At this point, static
analysis cannot determine which components will be connected at runtime and which
component the platform will choose, if there exists more than one suitable component.

In the end, a hybrid approach consisting of both static and dynamic analyses will be
reasonable.  This way, static analyses can be improved by information gained from
runtime analyses.  Furthermore, users who cannot afford to use static analysis tools
can rely on the TaintDroid approach, whereas in business or governmental scenarios as
well as at market entry, the static approach can be employed additionally.

\section{Related Work}

There exist a plethora of works for the static security analysis of software, e.g., discussed by Chess and West \cite{ChWe07}.  The
works on static information flow analysis for security often resides at the source code level. 
Some approaches deal with programmer-written annotations for
information flows that permit static code checks \cite{Myers1999,
Sabelfeld2003, Jif2010}. Moreover, the language-based security extensions in JFlow
\cite{Myers1999} support the modeling of access control. This allows one to statically
check code privileges, but all modeled access controls will be removed after the
JFlow compiler processing. This kind of language-based security analysis is limited
to the use of annotations by programmers at the source-code level. 
Our analysis methods, however, works without code annotations aiming to detect undesirable
information flow between different applications and components, respectively.

In another approach, type-based security combines annotations
with dependence graph-based information flow control \cite{Hammer2010}. Hammer's proposed analysis 
uses the Java bytecode and a succinct security policy specification that is
inserted as annotations in code comments. Although both approaches aim to detect information
flow violations of Java-based applications, they differ in the analysis techniques they use. 
We employ an analysis approach using the RFG to restrict the search space and
thereafter carrying out a more focused analysis at the
detail level. Hammer uses the complete dependence graph to directly conduct the information flow analysis. In addition, 
Hammer's method requires code annotations for the security labeling, similar to JFlow. This way, this approach can only be applied by the developer, but neither by the Android Market owner nor the end user.

Chandra and Franz implemented an information flow framework for Java applications using static as well as dynamic
checks \cite{conf/acsac/ChandraF07}. Static checks are needed to improve the dynamic analysis such that information 
about alternative control paths is also available. The approach works at the bytecode
level and is fully compatible with the class file
format. In general, all aforementioned approaches tackle the problem of 
indirect information flow induced by the control flow of applications, whereas we currently only analyze direct flows.
The focus of our work lies on an inter-application analysis. Furthermore, our techniques are tailored towards the Android platform by considering Android-specific programming concepts as well as data sources and sinks. Certainly, it is worthwhile to address indirect information flows in future work.

Important research prototypes from static security analysis are e.g.\ MOPS
\cite{Chen02}, Eau Claire \cite{Chess2002}, BLAST
\cite{Henzinger03}, and LAPSE \cite{Livshits05a}. MOPS uses temporal logics as formalism and model checking to
discover issues such as race conditions in C programs. Eau Claire allows the
formulation of pre- and postconditions as code annotations and is based on a theorem
prover.  Eau Claire detects general security problems such as buffer overflows. 
BLAST uses (lazy) abstraction to find safety properties in C/C++ code. 
The tool xg++ by Ashcraft and Engler was used to detect vulnerabilities in the 
Linux Kernel \cite{Ashcraft2002}. The LAPSE approach by Livshits and Lam resembles our approach
in also targeting Java applications. In contrast, we focus on interactions between applications and specifically consider Android's framework semantics for our analyses, whereas LAPSE aims to detect implementation bugs such as SQL injection and Cross-Site scripting vulnerabilities. However, the techniques employed by the LAPSE approach can also be used for our analyses at the source code level instead of the backward slicing algorithm. 

Some of the research prototypes have developed into commercial tools such as Fortify
Source Code Analyzer \cite{Fortify09} and Coverity Prevent \cite{Coverity10}. 
Our approach is complementary to all those works because we utilize
architectural information to focus the analysis at the code level. In addition, those
tools are designed to detect common low-level security bugs 
such as buffer overflows, SQL injection and Cross-site scripting vulnerabilities.
We, however, focus on information flow analysis, and more generally aim to detect 
security problems induced by software extensibility.

After the release of the Android platform, some works have addressed the built-in
application security mechanisms of this platform \cite{Enck2009,Shabtai2010}.
Moreover, Chaudhuri et al. define a formal language to describe
Android application behavior and the application's permission usage \cite{Chaudhuri2009}. 
As discussed above the TaintDroid approach is close to our work.
Beyond TaintDroid, there are also other approaches with similar goals. KIRIN is an alternative application installer for Android with a
built-in security framework that enforces policy invariants on the phone \cite{Enck2008}. The tool
checks at application install time for issues such as unchecked interfaces. When
problems are found, the application installation is canceled.  For the analysis, KIRIN only uses the
Android Manifest file (containing the permissions) and does not look at the program code. In addition, the interplay between
different applications was not considered. Nauman et al.\ present an Android
permission framework and a policy configuration user interface that allows the user to dynamically limit application
permissions at install time \cite{Nauman10}. Similarly to KIRIN, no
inter-component relations are taken into consideration for the policy
enforcement and only conditions according to time and usage count are described.
On a lower level, Shabtai et al.\
\cite{Shabtai09} enable the SELinux feature in the Android kernel to explore
additional protection opportunities and benchmark the system performance with
activated SELinux on a HTC G1 smartphone running Android.

Another research approach is the SAINT architecture \cite{Ongtang2009}. It inserts
enforcement hooks into Android's middleware layer to improve the currently limited
Android security architecture.  This work takes semantics such as location and time
into account, but strictly focuses on the developer view of permissions and does not
account for transitive data flows.

\section{Conclusion and Outlook}
\label{sec:summary}
In this paper, we discussed how the transitivity of trust problem affects
dynamic multi-component systems.  Focusing on the Android platform, we
presented a two-layer approach to the static security analysis of information flows
for composite Android applications and thus approached the transitivity of trust problem in this
context.  On the upper layer, we use the software architecture to
slice the applications. Thereafter, the actual data flow analysis is carried out at
the AST level. The results are integrated into the architecture to derive
information flows at the architectural layer.
We demonstrated the effectiveness of our analysis method with the
help of two real-world applications, which use advanced Java and Android programming concepts
such as inner classes, GUI handling, and Android service binding. 

There are several directions for further research.
First, we aim to support a more complete set of data sources and sinks as well as other concepts of the Android framework 
such as pending intents, URI permissions, and service hooks. 
Furthermore, our static analysis can be combined with dynamic analyses into a hybrid approach
in order to improve on the precision of the analyses. 
Lastly, we will analyze larger sets of applications. For example, it would be
interesting to investigate (at least) parts of the Android market and develop
information flow policies that the applications should adhere to.





\end{document}